\documentclass[reprint, aps,prl,twocolumn,showpacs,showkeys, 10pt, longbibliography, nolinenumbers, floatfix]{revtex4-2}
\usepackage{amsmath}
\usepackage{graphicx}
\usepackage{float}
\usepackage{dcolumn}
\usepackage{multirow}
\usepackage{natbib}
\usepackage[colorlinks = true,linkcolor = blue,urlcolor  = blue,citecolor = blue,anchorcolor = blue]{hyperref}
\usepackage{enumerate}
\usepackage{cleveref}

\begin{document}

\title{
Supramassive dark objects with neutron star origin}

\author{M. Vikiaris$^1$}

\author{V. Petousis$^2$}

\author{M. Veselsk\'y$^2$}

\author{Ch.C. Moustakidis$^1$}


\affiliation{$^1$Department of Theoretical Physics, Aristotle University of Thessaloniki, 54124 Thessaloniki, Greece\\
$^2$Institute of Experimental and Applied Physics, Czech Technical University }

\begin{abstract}
Till today, the nature of Dark Matter (DM) remains elusive despite all our efforts. This missing matter of the universe has not been observed by the already operating DM direct-detection experiments, but we can infer its gravitational effects. Galaxies and clusters of galaxies are most likely to contain DM trapped to their gravitational field. This leads us to the natural assumption that compact objects might contain DM too. Among the compact objects exist in galaxies, neutron stars are considered as natural laboratories, where theories can be tested, and observational data can be received. Thus, many models of DM have proposed it's presence in those stars. In particular, in the present study we focus on two types of dark matter particles, namely fermions and bosons with a mass range of [0.01-1.5] GeV  and self-interaction strength in the range [10$^{-4}$-10$^{-1}$] MeV$^{-1}$. By employing the two-fluid model, we discovered a stable area in the M-R diagram of a celestial formation consisting of neutron star matter and DM that is substantial in size. This formation spans hundreds of kilometers in diameter and possesses a mass equivalent to 100 or more times  the Solar mass. To elucidate, this entity resembles an enormous celestial body of DM, with a neutron star at its core. This implies that a supramassive stellar compact entity can exist without encountering any issues of stability and without undergoing a collapse into a black hole. In any case, the present theoretical prediction can, if combined with corresponding observations, shed light on the existence of DM and even more on its basic properties.

\keywords{Neutron stars, Dark matter, Two-fluid model,  Supramassive compact objects   }

\end{abstract}

\maketitle


\section{Introduction}
The mystery concerning the Dark Matter (DM) nature is still today one of the most challenging subjects in astrophysics and cosmology in general.  Two scenarios exist today, either DM is self-annihilating and through this procedure temperature gradient could be produced, or it is not self-annihilating and then the DM could cluster in blobs. But what is the possibility to have non self-annihilating DM captured and existing inside astrophysical compact objects?

We know that among the compact objects existing in galaxies, neutron stars are considered as natural laboratories, where theories can be tested, and observational data can be received~\cite{Shapiro-1983,Haensel-2007,Glendenning-2000}.
If it's possible that DM could clump in a sufficient  amount with nuclear matter in neutron stars,  this mixing could influence the neutron stars measurable properties (for a recent review see Ref.~\cite{Bramante-2024} and references therein). 
Thus, many models of DM have proposed it's presence in those stars either by studying how they accumulate in stars, how they affect their structure and basic properties or in more recent studies, how DM can be detected by measurements of tidal polarization in a binary system during the merger process~\cite{Goldman-89,Kouvaris-08,Bertone-08,Lavallaz-10,Kouvaris-10,Brito-15,Carmeno-17,Bramante-2013,Bell-2013,Leung-11,Leung-2012,Kouvaris-2012,Bertoni-13,Ciancarella-21,Panotopoulos-17,Panotopoulos-17b,Panatopoulos-2023,Das-20,Kumar-2020,Dasgupta-2021,Kouvaris-2011,Liu-2023,Dermott-2012,Hippert-2023,Goldman-2013,Dietl-2012,Li-2012,Xiang-2014,Guver-2014,Routaray-2023,AngLi-2012,Watts-2023,Raj-2018,Ivanytskyi-2020,Lopes-2023,Karkevandi-2022,Rezaei-2017,Sarkar-2020,Das-2021,Das-2022,Bell-2020,Seoane-2010,Shakeri-2022,Mariani-2024,Gresham-19, Ellis-18}. This suggests the exciting possibility that neutron stars could act as laboratories for indirectly measuring DM properties.  However, in the above works the study is limited to the case of the accretion of a small percentage of the DM so that its mass is also a small percentage of the total mass of the star. Another important question is how DM could become mixed with ordinary matter in a neutron star. One well studied possibility is through capture as described in the~\cite{Goldman-89,Kouvaris-08,Bertone-08,Lavallaz-10,Kouvaris-10,Brito-15,Carmeno-17,Bramante-2013,Bell-2013}. Also many other possibilities of how celestial bodies like neutron stars can accumulate DM and interact with neutron matter - nucleons investigated in previous studies~\cite{Thong-23, Sulagna-23, Maxim-19, Maxim-13}.  Relevant discussion about the nature, the self-interaction of DM and dark objects are presented in Refs.~\cite{Kaplan-09,Boucenna-14,Zurek-14,Kouvaris-15,Arkani-09,Navarro-96,Ray-2023,Maselli-2017,Nelson-19,Narain-06,Agnihotri-2009, Deliyergiyev-19,Zhang-2023,Freese-2016}. 

It's very useful to treat the mixing between nuclear matter and DM as a two-fluid systems, in which the first fluid describes the nuclear matter through an Equation of State (EoS) for a neutron star without DM and the second fluid describes the DM. 
The properties such as the mass and radius of the neutron star can be determined by solving the multi-fluid Tolman-Oppenheimer-Volkov (TOV) equations~\cite{Sandin-09,Kodama-72,Henriques-89}. 

Until today, no results were found from DM direct detection experiments~\cite{Akerib-17,Cui-17,Aprile-18} and thus have been placed strict constraints on the DM to the nucleons coupling strength. From the perspective of the TOV equations, this is generally taken to mean that the coupling strength its negligibly small and that DM admixed neutron stars are two-fluid systems in which the only inter-fluid interactions are gravitational~\cite{Gresham-19, Ellis-18}. 
Applying the two-fluid model in our study, we came to an surprising possibility of a stable area in the M-R diagram of a celestial formation consisting of neutron matter and DM that is substantial in size and vast in dimensions. This formation spans hundreds of kilometers in diameter and possesses a mass equivalent to 100 or more times  the Solar mass. To elucidate, this entity resembles an enormous celestial body of DM, with a neutron star at its core.  

This possibility implies that a supramassive stellar compact entity can exist without encountering any issues of stability and without undergoing a collapse into a black hole. It seems that the existence of such objects may demands mass of DM particle $m_{\chi}< 1~$GeV  or very strong self-interaction strength.   The lower is $m_{\chi}$ or/and stronger is the interaction, the higher is the $M_{max}$ and $R_{max}$. But before we claim something like this in our study, several crucial questions which they need convinced answers arise: 

a)  How stable are these objects?~\cite{Kain-21} b) What is their compactness? c)  Is there a consistency with the values of central pressure of the neutron star that do not exceed the accepted limits of the nuclear EoS? d) Is  the crucial assumption valid  for our investigation, that the fermions are stable on the time scale comparable with the lifetime of the universe, ($\tau\geq H_0^{-1}\approx 14$ Gyr, where $H_0$ is the value of the Hubble constant)? or in other words, holds the assumption that the fermions constituting the new compact object are conserved?~\cite{Narain-06} e) When and how the possible supramassive compact objects, made of exotic fermions, could be formed? One could speculate that this happened after the inflation era. Large objects could serve as primary seeds clumping nuclear matter after the  radiation decoupling era. Perhaps could be possible the formation of some hybrid objects where the exotic fermionic stars might be surrounded by the halo of ordinary matter. 

Answering most of the aforementioned questions, forms a large part of our research on the subject, especially the stability issue. Also in addition, the motivation behind our present work is to extend the recent study concerning the trapped DM in the interior of neutron stars and far beyond it.  It can be considered that the present study is a first  effort towards the  extension and generalization  of the study done in the papers~\cite{Narain-06,Agnihotri-2009}, but the main goal is to find  stable compact configurations of supramassive objects. In the aforementioned paper the study focused on compact stars consisting of one kind of fermions (or bosons) with arbitrary masses and interaction strengths. The present study include a similar study by considering a two-fluid model the one consist of neutron star matter and the other a DM fluid consist by fermions or bosons. In general, the calculation of the DM-EoS is a very difficult task due to the great uncertainty that exists in the knowledge about it and mainly concerns the mass and the interaction but also its concentration in the universe. Thus, only with some assumptions can one proceed with some predictions.  

In the present study we focus on the case of repulsive self-interaction between DM particles (whether fermions or bosons) based on known existing interactions. The repulsive interaction is fundamental to the stability of these objects, especially in the case of bosonic DM. The stronger the repulsive interaction, the greater the proportion of DM that can accumulate in these hybrid objects. It is fundamental for the stability of supramassive objects which is the object of study of this work. There are some justifications for strong repulsive interaction, such as some discrepancies at scales smaller than galaxy clusters could be addressed by a more complicated DM sector including large self-interactions (see Ref.~\cite{Nelson-19} and references therein). Moreover,  some recent work suggests that a strong velocity dependence in the interaction, such as would be the case if DM exchanges a light boson, provides the best fit to a range of galactic structures (for a relevant  discussion see Ref.~\cite{Nelson-19} and references therein).

We investigate the possibility to gain useful information and constraints of some possible DM candidates as well. Using a two-fluid model for neutron stars admixture with DM particles, we focus both on fermionic and bosonic DM where their mass and the self-interaction are treated as free parameters. Using this approach, we predict the properties of this new compact object including mass and radius. All of our findings are rigorously guided from the stability conditions of which these new objects must obey.

The paper is organized as follows: in Section 2, we briefly describe the two-fluid model used, accompanied with subsections dedicated to self interacting DM Equations of State, for fermionic and bosonic DM particles. In Section 3, we discuss the important topic of stability and in Section 4, we display and discuss the results of the present study. In Section 5, we finalize our investigation with the concluding remarks.




\section{Two-fluids model }
In the present work we study compacts objects made by a) non-self annihilating DM particles admixed with b) neutron star matter (including mainly neutrons and a small fraction of protons and electrons). We consider also that the total matter, described by these two fluids, interact each other only gravitationally. In order to predict the bulk properties of the objects consisting of the aforementioned two fluids,  one has to solve the coupled Tolman-Oppenheimer-Volkov (TOV) equations (two for each fluid) simultaneously. These four equations are defined below~\cite{Kodama-72,Sandin-09}:       
\begin{eqnarray}
\frac{dP_{\rm NS}(r)}{dr}&=&-\frac{G{\cal E}_{\rm NS}(r) M(r)}{c^2r^2}\left(1+\frac{P_{\rm NS}(r)}{{\cal E}_{\rm NS}(r)}\right) \nonumber\\
&\times&
 \left(1+\frac{4\pi P(r) r^3}{M(r)c^2}\right) \left(1-\frac{2GM(r)}{c^2r}\right)^{-1}
\label{TOV-1}
\end{eqnarray}
\begin{equation}
\frac{dM_{\rm NS}(r)}{dr}=\frac{4\pi r^2}{c^2}{\cal E}_{\rm NS}(r)
\label{TOV-2}
\end{equation}
\begin{eqnarray}
\frac{dP_{\rm DM}(r)}{dr}&=&-\frac{G{\cal E}_{\rm DM}(r) M(r)}{c^2r^2}\left(1+\frac{P_{\rm DM}(r)}{{\cal E}_{\rm DM}(r)}\right) \nonumber \\
&\times&
 \left(1+\frac{4\pi P(r) r^3}{M(r)c^2}\right) \left(1-\frac{2GM(r)}{c^2r}\right)^{-1}
\label{TOV-1}
\end{eqnarray}
\begin{equation}
\frac{dM_{\rm DM}(r)}{dr}=\frac{4\pi r^2}{c^2}{\cal E}_{\rm DM}(r)
\label{TOV-2}
\end{equation}
where also $M(r)=M_{\rm NS}(r)+M_{\rm DM}(r)$ and $P(r)=P_{\rm NS}(r)+P_{\rm DM}(r)$ (the subscripts $\rm NS$ and $\rm DM$ stand for the neutron star and dark matter respectively).

\section{Neutron star matter and  dark matter equations of state}

In  the present work, in order to describe the neutron star matter we use the Equation of State (EoS) derived by  Akmal et al.~\cite{Akmal-1998} (hereafter APR) in particular the model  A18+UIX. This is an EoS with microscopic origin and its predictions are in very good agreement with both the measured maximum masses (see PSR J1614-2230~\cite{Arzoumanian-2018}, PSR J0348+0432~\cite{Antoniadis-2013}, PSR J0740+6620~\cite{Cromartie-2020}, 
and PSR J0952-0607~\cite{Romani-2022} pulsar observations for the possible maximum mass) and some astrophysical constraints for radii (see the GW170817 event~\cite{Abbott-2019-X}). It is worth to notice here that the present study mainly focuses on the EoS of DM and its properties. Nevertheless, the use of a realistic EoS for neutron stars provides a guarantee for the reliability of our results.
 

Now, concerning the DM particles we consider that they are relativistic fermions, which interact each other through a repulsive force. In particular, we consider Yukawa type interaction on the form~\cite{Nelson-19}: 
\begin{equation}
V(r)=\frac{{\rm g}_{\chi}^2 (\hbar c)}{4\pi r} \exp\left[-\frac{m_{\phi}c^2}{\hbar c}r\right]
\label{Yukawa-1}
\end{equation}
where 
${\rm g}_{\chi}$ is  the coupling constant  and $m_{\phi}$ is the mediator mass. It is worth to mention that in the majority of the similar studies, the DM considered as a non-interacting gas of fermions. However, as we will see in the present study, the role of interaction is decisive for the creation and stability of supramassive objects.
The contribution on the energy density of the self-interaction is given by: 
\begin{equation}
{\cal E}_{SI}(n_{\chi})=\frac{y^2}{2} (\hbar c)^3 n_{\chi}^2  
\end{equation}
where $y={\rm g}_{\chi}/m_{\phi}c^2$ (in units MeV$^{-1}$).
In this case the  total energy density of the DM particles is given by~\cite{Narain-06}:
\begin{eqnarray}
{\cal E}_{\rm DM}(n_{\chi})&=&\frac{(m_{\chi}c^2)^4}{(\hbar c)^38\pi^2}\left[x\sqrt{1+x^2}(1+2x^2) \right. \nonumber\\
&-&\left. \ln(x+\sqrt{1+x^2})\right] + \frac{y^2}{2} (\hbar c)^3 n_{\chi}^2
\label{Rel-ED}
\end{eqnarray} 
The corresponding  pressure is calculated straightforward by using the definition:
\begin{equation}
P_{\rm DM}(n_{\chi})=n_{\chi}\frac{d {\cal E}_{\rm DM}(n_{\chi})}{d n_{\chi}}-{\cal E}_{\rm DM}(n_{\chi})
\label{chi}
\end{equation}
and is given by: 
\begin{eqnarray}
P_{\rm DM}(n_{\chi})&=&\frac{(m_{\chi}c^2)^4}{(\hbar c)^38\pi^2}\left[x\sqrt{1+x^2}(2x^2/3-1)
\right. \nonumber \\
&+&\left. \ln(x+\sqrt{1+x^2})\right]  + \frac{y^2}{2} (\hbar c)^3 n_{\chi}^2
\label{Rel-Pr}
\end{eqnarray}
where $m_{\chi}$ is the particle mass and 
\[x=\frac{(\hbar c)(3\pi^2n_{\chi})^{1/3}}{m_{\chi}c^2}.  \]
The total energy density and pressure of the two-fluid mixing, considering that the contribution of neutron star matter are defined as ${\cal E}_{\rm NS}(n_b)$  and $P_{\rm NS}(n_b)$ where both are functions of the baryon density $n_b$ (for more details see Ref.~\cite{Akmal-1998}), are given simply by the following sums: 
\begin{equation}
{\cal E}(n_b,n_{\chi})={\cal E}_{\rm NS}(n_b)+{\cal E}_{\rm DM}(n_{\chi})
\label{eos-1-kouv}
\end{equation}
\begin{equation}
P(n_b,n_{\chi})=P_{\rm NS}(n_b)+P_{\rm DM}(n_{\chi})
\label{P-tot-1}
\end{equation}

With the aim of enriching the possible DM candidates, we will also consider the case where it consists of bosonic particles. We use in this case the EoS provided in Ref.~\cite{Agnihotri-2009} and used recently in Ref.~\cite{Watts-2023}, where the energy density and pressure are given respectively by: 
\begin{eqnarray}
{\cal E}_{\rm DM}(n_{\chi})&=&m_{\chi}c^2n_{\chi}+\frac{u^2}{2} (\hbar c)^3 n_{\chi}^2,\nonumber \\
P_{\rm DM}(n_{\chi})&=& \frac{u^2}{2} (\hbar c)^3 n_{\chi}^2
\label{Bos-3}
\end{eqnarray}
where the quantity $u={\rm g}_{\chi}/m_{\phi}c^2$ (in units MeV$^{-1}$) defines the strength of the interaction in analogy with the case of fermions.  It is worth emphasizing here, that as it was  found firstly in Ref.~\cite{Colpi-1986}, the structure of the boson stars (mass and radius), even for low values of the self-interaction, can differ radically, compared to the case of non-interacting. Then, by suitable choice of the interaction, the masses of the corresponding boson stars could be comparable to the Chandrasekhar mass of fermion stars. These boson stars could arise during the gravitational condensation of bosonic DM in the early Universe~\cite{Colpi-1986}.

\section{Stability}
The most usual method to study the stability  is to consider small radial perturbations of the
equilibrium configuration by solving the Sturm-Liouville
eigenvalue equation, which yields eigenfrequencies $\omega_n$~\cite{Shapiro-1983}. The eigenfrequencies of the different modes form a
discrete hierarchy $\omega_n^2<  \omega_{n+1}^2$
$n=0,1,2\cdots$ where $\omega_n$ being real
numbers~\cite{Dengler-2022}. A negative value of $\omega_n^2$
leads to an exponential
growth of the radial perturbation and collapse of the star. Only
when all eigenfrequencies are positive the star will be
stable~\cite{Bardeen-1966} (see also the relevant discussion in Ref~\cite{Dengler-2022}).
A relevant study of pulsation equations for compact stars made up of an arbitrary number of perfect fluids has been developed recently in Ref.~\cite{Kain-2020}.

In the present work we employ the method developed by Henriques, Liddle and Moorhouse which was used for the study of boson-fermion star~\cite{Henriques-1989,Henriques-1990a,Henriques-1990b}. This method has been elaborated and extended through  the years for similar 
studies~\cite{Kain-21,Alvarado-2013,Alvarado-2020,Giovanni-2022}.  According to this method, the stability analysis is carried out by examining the behaviour of baryons and DM particles, yet fixing the total mass $M$ values. In particular, the stability curve is formed with the pair of central values of pressures $\{P_c^{\rm NS}, P_c^{\rm DM}$\} exactly in the point where the number of particles reached the minimum and maximum values. This critical curves identify the transition from linear stable and unstable with respect to perturbations that conserve the mass and the particle number. Hence fulfill the following conditions~\cite{Giovanni-2022}:
\begin{eqnarray}
&&\left(\frac{\partial N_b}{\partial P_c^{\rm NS}}\right)_{\rm M=const}=\left(\frac{\partial N_{\chi}}{\partial P_c^{\rm NS}}\right)_{\rm M=const}=0 \nonumber \\
&&\left(\frac{\partial N_b}{\partial P_c^{\rm DM}}\right)_{\rm M=const}=\left(\frac{\partial N_{\chi}}{\partial P_c^{\rm DM}}\right)_{\rm M=const}=0
\end{eqnarray}
where $N_b$ and $N_{\chi}$ are the numbers of baryons and DM particles respectively. These
numbers are given by the 
expression~\cite{Gresham-19}
\begin{equation}
N_{i}=4\pi\int_{0}^{R_i} n_i(r) \frac{r^2 dr}{\sqrt{1-2GM(r)/rc^2}}, \quad i=b,\chi
\end{equation}
where $R_i$ is the radius of the corresponding fluid. 
Now, the stability analysis can be summarized as follows~\cite{Giovanni-2022}: Sequences of stable equilibrium configurations that start from purely DM star have the feature that the number of DM particle $N_{\chi}$ first decreases (as a function of the central pressures $P_c^{\rm DM}$ or $P_c^{\rm NS}$) down to the critical point and the number of baryons $N_b$ increases. In general, equilibrium sequences can consist of continuous regions of stability and instability. In this case, therefore, we can have more than one branch of stability.


\section{Results and Discussion}
What should be emphasized first is the basic difference between finding the stability of the two-fluid model and that of the one-fluid model. Thus, while in the one-fluid model, the finding of stability is limited to a curve on an M-R diagram, in the two-fluid model the curve has essentially been replaced by a surface on a corresponding diagram. This is due to the use of three independent parameters, that is the DM mass $m_{\chi}$, the strength of the self-interaction and the fraction $f=P_{DM}^c/P_{NS}^c$ of the central densities. Basically, all of them are unconstrained. In the present study we focus on a few stable configurations  which prove and confirm our hypothesis for the existence of supramassive dark objects in the universe, with neutron star origin. 

It is proper to clarify here that, by supramassive compact objects, we mean those whose compactness $C=GM/Rc^2$ is comparable to those of  neutron stars, i.e. $C\in $ [0.05-0.30].  In any case, the effects of the  general relativity are important and moreover the curvature of spacetime in the neighborhood of these objects is noticeable. This has the consequence that they can be perceived in this way.  

It is worthwhile to notice that the majority of the relevant studies dedicated to the specific cases, where the hybrid objects consist mainly by hadronic matter and a small fraction of DM, leading to configurations with mass similar to neutron stars but with DM halo extended even to a few tens of kilometers. 

To be more specific, in the present study we investigated the possibility of very massive hybrid objects with masses ranging from ten to even hundred solar masses and radii also for tens to even thousand of kilometers. In principle there is no ad-hoc limitation to these configurations except for the stability condition. Some additional constraints can be introduced by the absorption rate of DM (in particular of the DM particle-neutron scattering cross section), in a neutron star and the associated time scales, or by the density distribution of DM in the universe. Besides, compact DM objects with large masses and corresponding radii have already been studied (for a systematic review see references~\cite{Narain-06,Agnihotri-2009}. The motivation of the present work is to study the corresponding supramassive compact objects, where a neutron star sits at their center triggering the accretion of DM in and around it.

In general, as already pointed out, calculating the stability of these objects is a computationally difficult and time-consuming problem. Since we are initially interested in proving the stability of these objects we will choose some special cases whose stability we will confirm. A more systematic and extensive study may determine the stable range (concerning the mass and radius), of such objects as a function of the mass,  interaction and the DM contribution  to the central pressure.

Now, since the DM contributes as larger fraction on the total mass, some useful approximation, concerning the total number of fermions in the case  of the maximum mass configurations, can be inferred by applying the analytical solution of the TOV equations, that is the Tolman VII solution~\cite{Tolman-39}. According to this solution, the energy density varies in the interior of the star according the rule ${\cal E}(r)\sim(1-(r/R)^2)$. In this case the binding energy, $BE/c^2=N_{\chi}m_x-M$, is given by~\cite{Lattimer-2005}
\begin{equation}
\frac{BE}{Mc^2}=f(C), \quad f(C)=\frac{11 }{21}C+\frac{7187}{18018}C^2+\cdots 
\end{equation}
where $f(C)$ is a function of the compactness of the star $C=GM/Rc^2$. After some algebra, we get the following approximation, for the critical number of fermions which corresponds to the stability limit: 
\begin{equation}
 N_{\chi}^{\rm cr}=1.144\times 10^{51} \left(0.627+0.27\tilde{y}\right)\left(1+f(C_{\rm cr})\right) \left(\frac{100 \ {\rm GeV}}{m_{\chi}} \right)^3 
 \label{Nx-approx}
\end{equation}   
where $C_{\rm cr}=GM_{\rm max}/R_{\rm min}c^2$ (for a relevant discussion see also Ref.~\cite{Maselli-2017}). 
The values $M_{\rm max} $ and $R_{\rm min}$  are taken from  Ref.~\cite{Narain-06} where: 
\begin{equation}
 M_{\rm max } =\left(0.384+0.165 \tilde{y}  \right)\times \left(\frac{\rm GeV}{m_{\chi}}  \right)^2\times 1.632 \ M_{\odot}  
\end{equation}
\begin{equation}
 R_{\rm min  } =\left(3.367+0.797 \tilde{y}  \right)\times \left(\frac{\rm GeV}{m_{\chi}}  \right)^2\times 2.410 \ ({\rm km})  
\end{equation}
where $\tilde{y}$, a dimensionless quantity, related with the interaction which runs in the range $10^{-2}$ to $10^3$~\cite{Narain-06} and it is related with the parameter $y$ according to $\tilde{y}=(m_{\chi}c^2/\sqrt{2})y$.

\begin{figure}[h]
\centering  
\includegraphics[width=70mm, height=65mm]{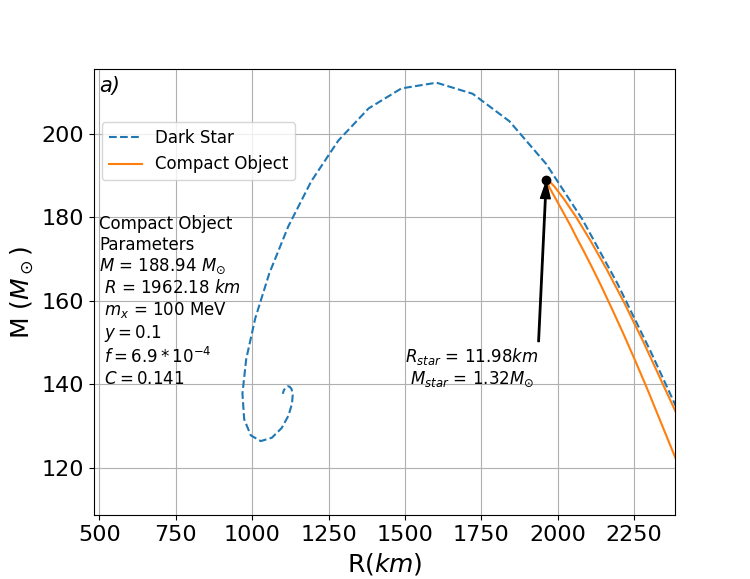}\\
\includegraphics[width=70mm, height=65mm]{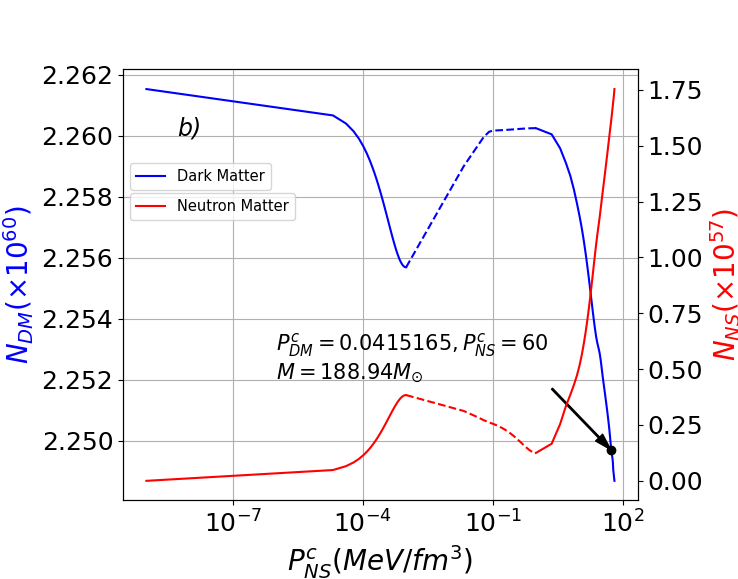}
\caption{(a) The M-R dependence for the case of pure fermionic DM (dashed line) and compact object (solid line) of DM fermion with mass $m_{\chi}=100$ MeV (the rest properties, that is the interaction strength $y$ (in MeV$^{-1}$), the fraction $f$ and the compactness $C$ also displayed in the figure).   The arrow   indicates  the specific case of a compact object with   $M=188.94\  M_{\odot}$ and $R=1962.18$ km  (correspond to  a neutron star with mass  $M=1.32 \ M_{\odot}$  and radius $R=11.98$ km  in the center). (b) 
The dependence of the number of DM particles $N_{\rm DM}$ and baryons $N_{\rm NS}$  on  the pressure    $P_{\rm NS}^c$  for equlibrium configurations of equal mass $M=188.94 \ M_{\odot}$. 
The solid lines indicate the stable branches, while the dashed lines are the unstable ones.  The arrow corresponds to the specific  stable configuration which indicated also by an arrow in the top panel (a).}
\label{M-R-Ferm-100-05}
\end{figure}

\begin{figure}[h]
\centering  
\includegraphics[width=70mm, height=65mm]{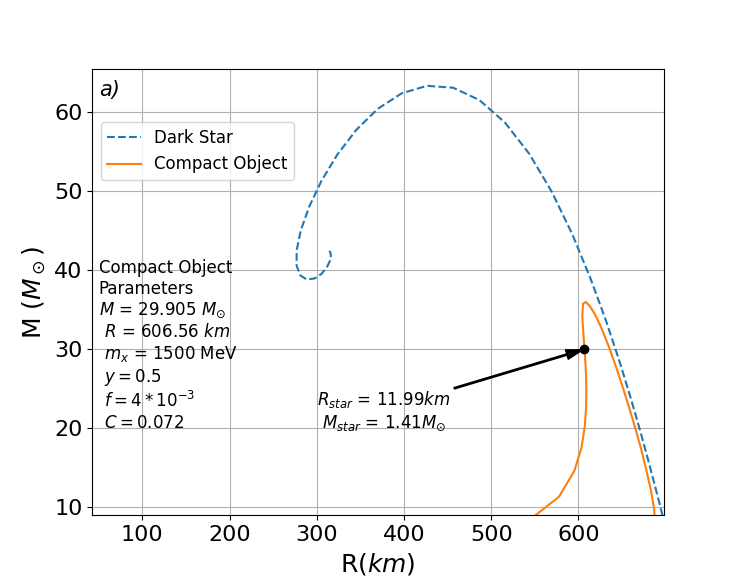}\\
\includegraphics[width=70mm, height=65mm]{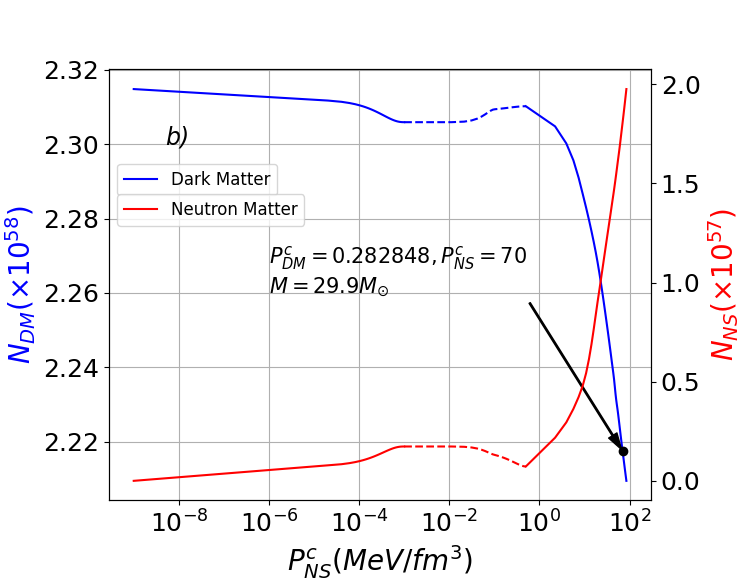}\\ 
\caption{(a) The same as in Fig.~\ref{M-R-Ferm-100-05}(a) for a DM fermion with mass $m_{\chi}=1500$ MeV.  We  indicate the specific  case $M=29.905 \ M_{\odot}$ and $R=606.56$ km, which include  a neutron star in the center  with mass  $M=1.41 \ M_{\odot}$  and $R=11.99$ km. (b) The same as in Fig.~\ref{M-R-Ferm-100-05}(b).  }
\label{M-R-1500-Ferm-0.5}
\end{figure}

The expression (\ref{Nx-approx}) is similar to that found in Ref.~\cite{Dermott-2012} (using a Newtonian approach) 
where:
\begin{equation}
N_{\chi}^{\rm cr}\simeq 1.8 \times 10^{51}\left(\frac{100 \ {\rm GeV}}{m_{\chi}} \right)^3
\label{N-ch-ferm}
\end{equation}

When the number of DM particles, concentrated in a neutron stars exceed the Chandrasekhar limit the compact object collapse into a black hole \cite{Dermott-2012}. It is concluded then that the observations of old neutron stars can be used as a laboratory to constrain the DM - neutron scattering cross section~\cite{Dermott-2012}.

In Fig.~\ref{M-R-Ferm-100-05}(a), we display the M-R dependence for the case of pure fermionic DM object and DM-neutron star mixture object (indicated as compact object) for $m_{\chi}=100$ MeV,  $y=0.1$ (in MeV$^{-1}$) and fraction $f=6.9\cdot 10^{-4}$. Obviously there is an infinite number of possible configurations but we focus on the indicated case $M=188.94 \ M_{\odot}$ and $R=1962.18$ km with a neutron star $M=1.32 \ M_{\odot}$  and $R=11.98$ km  in the center. This  object, with  compactness $C=0.141$, indicated with an arrow. In Fig.~\ref{M-R-Ferm-100-05}(b), we confirm the stability of this object, with the help of the $P_{\rm NS}^c-N_{\rm DM, NS}$ diagram. The object, which is indicated also with the arrow, belongs to the second stable branch of the $P_{\rm NS}^c-N_{\rm DM, NS}$ diagram.

\begin{figure}[h]
\centering  
\includegraphics[width=70mm, height=65mm]{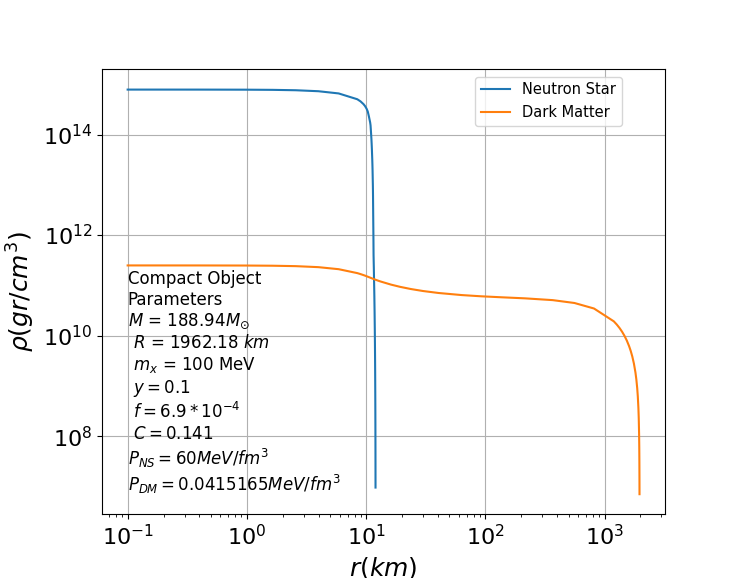}\\
\caption{The density of DM and neutron star matter as a function of the distance $r$ for the specific configuration denoted in Fig.~\ref{M-R-Ferm-100-05}(a). }
\label{RDM-PNS-r}
\end{figure}


\begin{figure}[h]
\centering  
\includegraphics[width=70mm, height=65mm]{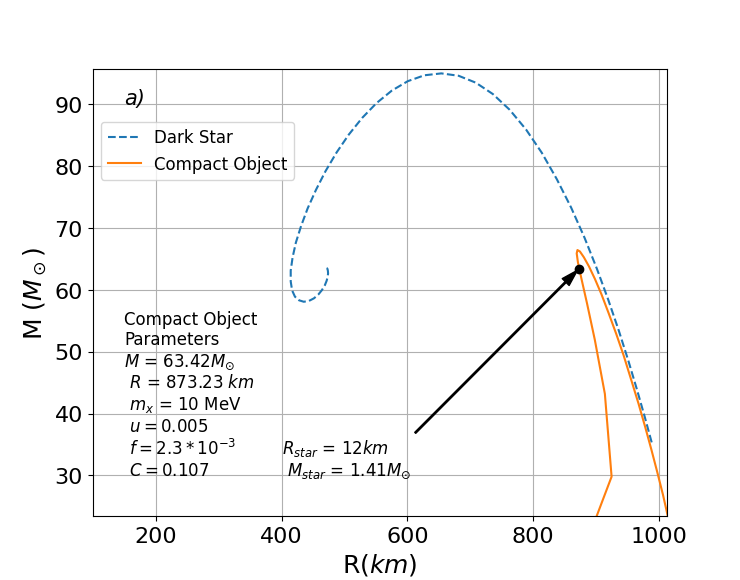}\\
\includegraphics[width=70mm, height=65mm]{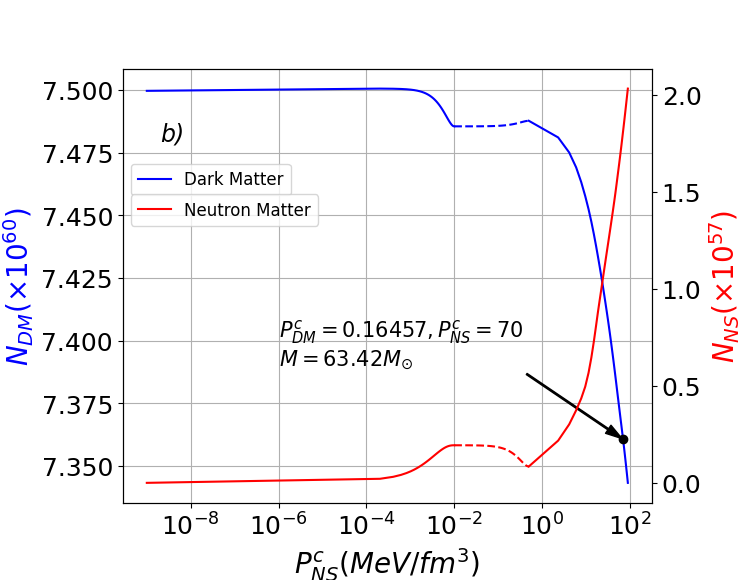}\\ 
\caption{(a) The M-R dependence for the case of pure bosonic DM (dashed line) and compact  object (solid line) of a DM boson with mass  $m_{\chi}=10$ MeV. The arrow indicate the specific case where  $M=63.42 \  M_{\odot}$ and $R=873.23$ km  (correspond to  a neutron star with mass  $M=1.41 \  M_{\odot}$  and radius $R=12$ km). (b) The same as in Fig.~\ref{M-R-Ferm-100-05}(b).   }
\label{M-R-Boson-10-0.5}
\end{figure}

\begin{figure}[h]
\centering  
\includegraphics[width=70mm, height=65mm]{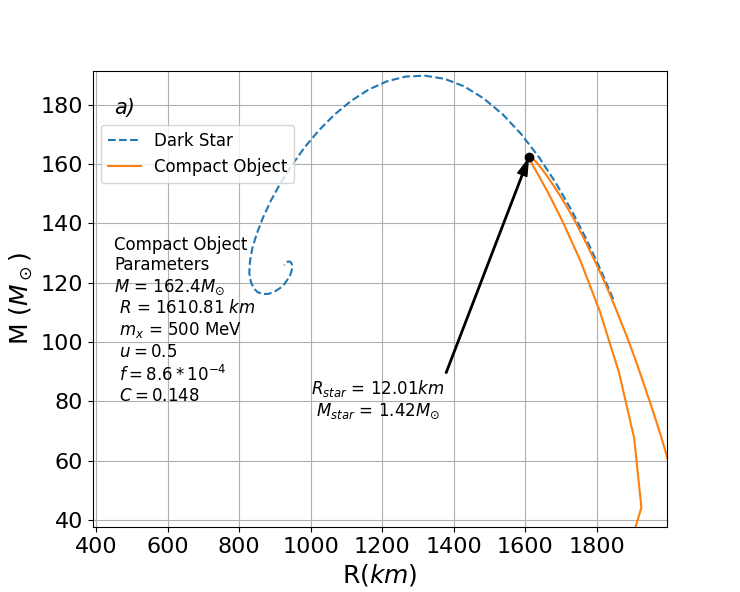}\\
\includegraphics[width=70mm, height=65mm]{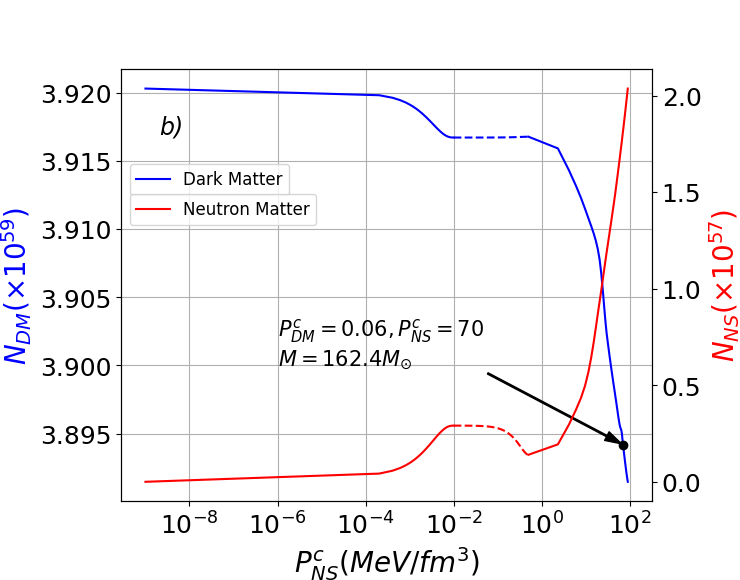}\\ 
\caption{ (a) The same as in Fig~\ref{M-R-Boson-10-0.5}(a) for a  DM boson with mass  $m_{\chi}=500$ MeV.   We indicate the specific case where  $M=162.4 \ M_{\odot}$ and $R=1610.81$ km  (correspond to  a neutron star with mass  $M=1.42 \ M_{\odot}$  and radius $R=12.01$ km). (b)  The same as in Fig.~\ref{M-R-Ferm-100-05}(b). }
\label{M-R-Boson-100-0.5}
\end{figure}

In Fig.~\ref{M-R-1500-Ferm-0.5}(a), we investigate the case of a heavier DM particle with mass  $m_{\chi}=1500$ MeV.  We  select  the indicated case $M=29.905 \ M_{\odot}$ and $R=606.56$ km with a neutron star $M=1.41 \ M_{\odot}$  and $R=11.99$ km  in the center. Again,  in Fig.~\ref{M-R-1500-Ferm-0.5}(b), we confirm the stability of this object.  

Moreover, in Fig.~\ref{RDM-PNS-r} we display the dependence of the pressure (which corresponds to neutron star matter and DM) on the radius $r$ for the specific configuration denoted in Fig.~\ref{M-R-1500-Ferm-0.5}(a).

In Fig.~\ref{M-R-Boson-10-0.5}(a), we consider the case of a bosonic DM particle with mass $m_{\chi}=10$ MeV. We select the specific configuration which corresponds to total mass  
$M=63.42 \ M_{\odot}$, $R=873.23$ km (with a neutron star with mass  $M=1.41 \ M_{\odot}$  and radius $R=12$ km). This compact object is stable as one can see in Fig.\ref{M-R-Boson-10-0.5}(b).

In Fig.~\ref{M-R-Boson-100-0.5}(a)  we select the case  of a  DM boson with mass $m_{\chi}=500$ MeV. We select the configuration  where  $M=162.4 \ M_{\odot}$, $R=160.81$ km   (with  a neutron star with mass  $M=1.42 \ M_{\odot}$  and radius $R=12.01$ km). This object is stable as confirmed in Fig.~\ref{M-R-Boson-100-0.5}(b).   

\begin{figure}[h]
\centering  
\includegraphics[width=70mm, height=65mm]{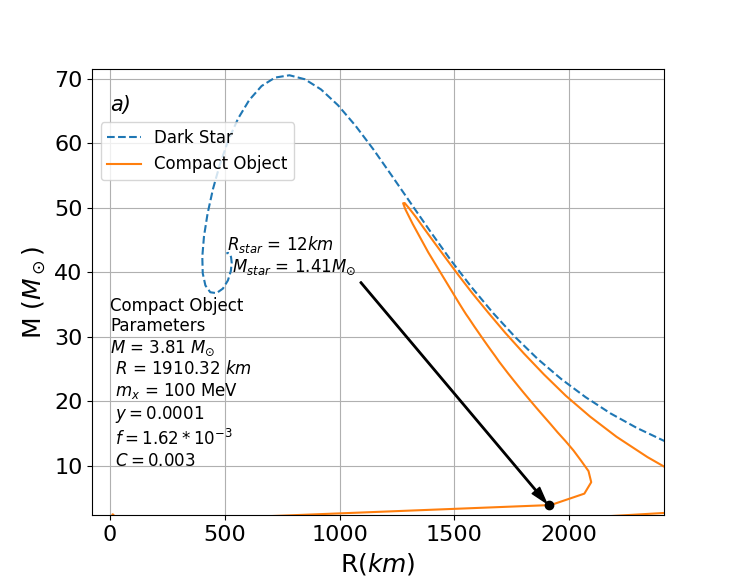}\\
\includegraphics[width=70mm, height=65mm]{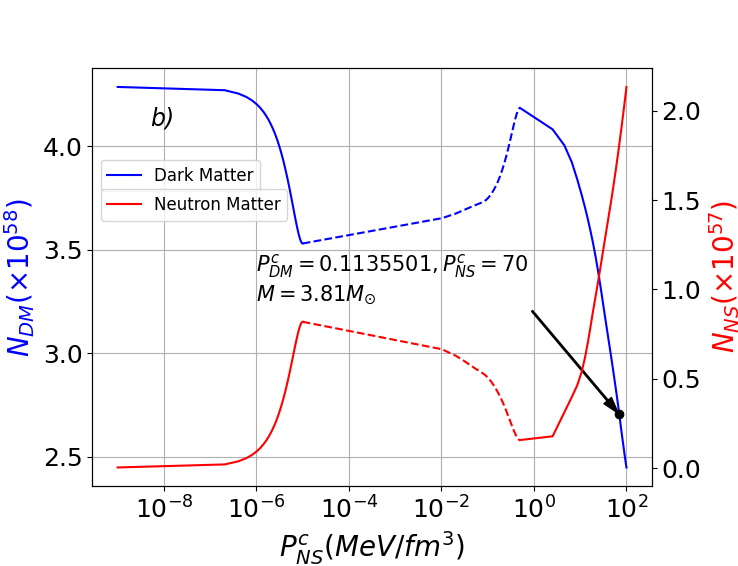}\\ 
\caption{ (a) The same as in Fig.~\ref{M-R-Ferm-100-05}(a) for a DM fermion with mass $m_{\chi}=100$ MeV.  We  indicate the specific  case $M=3.81 \  M_{\odot}$ and $R=1910.32$ km, which include  a neutron star in the center  with mass  $M=1.41 \ M_{\odot}$  and $R=12$ km. (b) The same as in Fig.~\ref{M-R-Ferm-100-05}(b).   }
\label{M-R-100-mas-gap}
\end{figure}

In the previous cases we focused on the case of supramassive  objects. However, the case of a compact object in the region of the mass gap between neutron stars and black holes is also of great interest in Astrophysics (for a recent discussion see Ref.~\cite{Fishbach-2024}). According to the findings of the present paper, there may be a stable configuration of objects with mixing of DM with neutron stars in this region. To be more specific,  we display in Fig.~\ref{M-R-100-mas-gap}(a) the M-R dependence for the case of pure fermionic DM object and DM-neutron star mixture object (indicated as compact object) for $m_{\chi}=100$ MeV,  $y=0.0001$ and fraction $f=1.6\cdot 10^{-3}$.
We focus on the indicated case $M=3.81 \ M_{\odot}$ and $R=1910.32$ km with a neutron star $M=1.41 \ M_{\odot}$  and $R=12$ km  in the center. This is an  objects with compactness $C=0.003$.  In Fig.~\ref{M-R-100-mas-gap}(b), we confirm the stability of this object, with the help of $P_{\rm NS}^c-N_{\rm DM, NS}$ diagram.

\begin{figure}[h]
\centering  
\includegraphics[width=70mm, height=65mm]{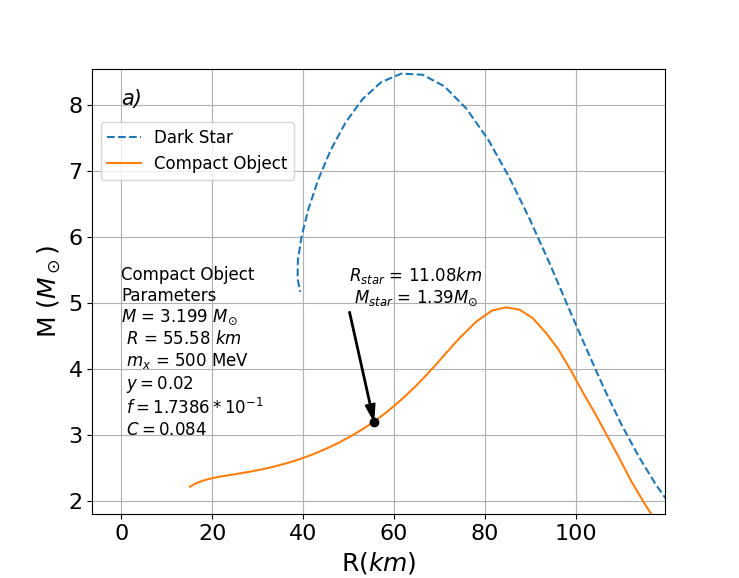}\\
\includegraphics[width=70mm, height=65mm]{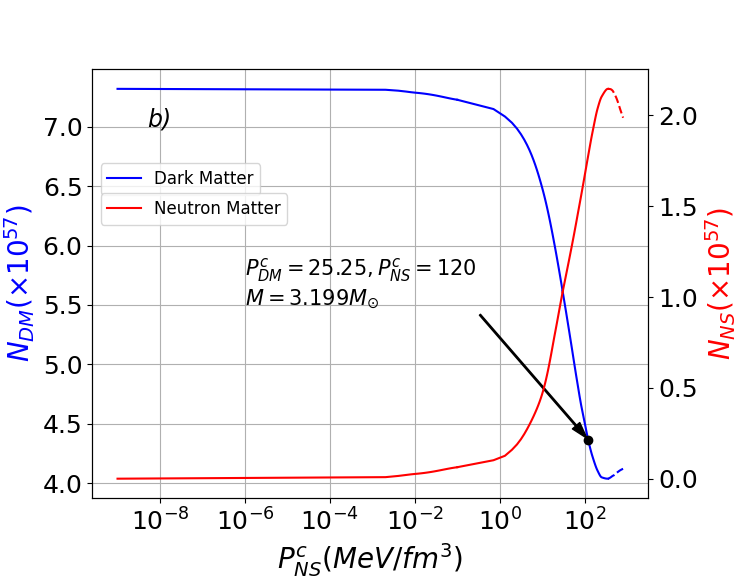}\\ 
\caption{ (a) The same as in Fig.~\ref{M-R-Ferm-100-05}(a) for a DM fermion with mass $m_{\chi}=500$ MeV.  We  indicate the specific  case $M=3.199 \ M_{\odot}$ and $R=55.58$ km, which include  a neutron star in the center  with mass  $M=1.39 \ M_{\odot}$  and $R=11.08$ km. (b) The same as in Fig.~\ref{M-R-Ferm-100-05}(b).   }
\label{M-R-500-mas-gap}
\end{figure}

Finally, in Fig.~\ref{M-R-500-mas-gap}(a), we investigate the case of a heavier DM particle with mass  $m_{\chi}=500$ MeV.  We  select  the indicated case $M=3.199 \ M_{\odot}$ and $R=55.58$ km with a neutron star $M=1.39 \ M_{\odot}$  and $R=11.08$ km  in the center. Again,  in Fig.~\ref{M-R-500-mas-gap}(b), we confirm the stability of this object. 

From the above study it is clear that the existence of objects in the mass-gap region can be justified if one takes into account a suitable combination of DM parameters. These objects may resemble neutron stars but have a different structure and composition.

\section{Concluding Remarks}
We investigated possible supramassive dark objects, stable configurations with neutron star origin. We used EoS of self-interacting fermions and bosons and one of the most reliable EoS for the neutron star matter. Unlike other similar studies, we did not focus on the cases of neutron stars surrounded by a DM halo of a few kilometers and of low relative mass contribution of the DM. We found that the number of these configurations is unlimited (just we select few of them to present), which are also stable against perturbations. This means that they may exist in the universe. An extension of this study is to include interactions between DM particles and baryons. This is an interesting perspective to be studied in future work.

In the  present work we do not indent to answer on the question "how these objects are created?". There are few speculations related with this question. A relevant discussion is provided in Ref.~\cite{Dengler-2022}. For example the accretion mechanism of DM into neutron stars as by the primordial formation of DM clumps surrounded by ordinary matter. Another possibility is of having dark compact object originated from DM perturbations growing from primordial over-densities. Finally, another possibility is copious production and capture of DM in the core-collapse supernova of neutron stars progenitor~\cite{Gresham-19}. In any case, there is not any strong theoretical prediction which argues strongly against the creation of these supramassive DM objects. Some other works~\cite{Sulagna-23} also suggesting that it could be possible,
without assuming any priors on DM parameters, to have gravitational waves 
detection of non-annihilating heavy DM. This detection will  cover the entire
parameter space that explains the missing pulsars, and
going well below the neutrino floor.  

The last proposition brings us to the most critical question, how we could locate them (as far as the objects are concerned) ? The most powerful tool in this case is the gravitational lensing which is based on space-time distortion which is caused by these objects. Another possibility is to detect a possible merger event, by the well known detectors (LIGO-VIRGO-KAGRA collaboration) which include two cases: a) merger between two dark compact objects or b) merger between one of them and another compact object including pure neutron star or  black hole. The  signal of the corresponding gravitational waves will transfer useful information for the structure of these objects~\cite{Bauswein-2023}. Another possible scenario to detect these objects, which was proposed in Ref.~\cite{Giovanni-2022}, could be that of a dark star in the second branch that accreted a low amount  of neutron star matter and that lost part of the bosonic matter due to accretion onto a second more compact object, which lead the star  to the first unstable branch and triggering the dispersion mechanism~\cite{Giovanni-2022}. 

In Ref.~\cite{Freese-2016} the authors present in a systematic way the creation of giant dark stars in a baryonic mass environment. In particular,  they claim that since the dark stars reside in a large reservoir of baryons, the baryons can start to accrete onto the dark stars. The evolution of dark stars start from their inception at $~ 1 M_{\odot}$ and and as they accumulated mass from
environment grow to become supermassive stars, possibly as large as $10^7 M_{\odot} $~\cite{Freese-2016}. Finally, as pointed out in Ref.~\cite{Bramante-2024} (for more details see also references therein) exotic compact objects with  near-Schwarzschild compactness may exist and constitute DM, e.g., some classes of boson stars, Q-balls, non-topological solitons, and ultra-compact minihalos.  

Moreover, in the present work we explored the  interesting possibility that some of recent measurement's of very massive neutron stars~\cite{Abbott-2020,Ligo-2024,Barr-2024} and  stars in the region of  black hole-neutron stars mass gap can be attributed to the capture of DM by neutron stars. In any case, further theoretical studies and precise astrophysical observations will help  to shed light on this open issue.   


The presented study was the first attempt of investigating supramassive compacts objects with neutron star origin, guided by the stability criterion. We intend to extend our investigation focusing on the mass gap region and to perform a categorization of our studies. 


\section*{Acknowledgments}
This work is supported by the Czech Science Foundation (GACR Contract  No. 21-24281S).




\end{document}